# Symmetries and alignment of biaxial nematic liquid crystals

Panagiota K. Karahaliou Alexandros G. Vanakaras and Demetri J. Photinos
*Department of Materials Science, University of Patras, Patras 26504, Greece*

The possible symmetries of the biaxial nematic phase are examined against the implications of the presently available experimental results. Contrary to the widespread notion that biaxial nematics have orthorhombic symmetry, our study shows that a monoclinic ($C_{2h}$) symmetry is more likely to be the case for the recently observed phase biaxiality in thermotropic bent-core and calamitc tetrapode nematic systems. The methodology for differentiating between the possible symmetries of the biaxial nematic phase by *NMR* and by *IR* spectroscopy measurements is presented in detail. The manifestations of the different symmetries on the alignment of the biaxial phase are identified and their implications on the measurement and quantification of biaxiality as well as on the potential use of biaxial nematic liquid crystals in electro-optic applications are discussed.

## I. INTRODUCTION

The structurally simplest liquid crystals are the uniaxial, a-chiral, a-polar nematics.[1, 2] They are positionally uniform and inversion-symmetric anisotropic media that have a unique axis of full rotational symmetry, referred to as the director **n**. The corresponding liquid crystal phase, often denoted as $N_u$ to distinguish it from other types of nematic phases, has the symmetries of the $D_{\infty h}$ point group. The anisotropy of the $N_u$ phase is primarily reflected on second-rank tensor physical properties and is quantified by means of second-rank tensor orientational order parameters. Nematics of lower symmetry are in principle conceivable on relaxing any one of the three basic symmetries of the $N_u$ phase_ i.e. a-chirality, a-polarity and full rotational symmetry about the director_ or any combination thereof. However, of all the conceivable lower symmetry nematics, only the chiral ones ($N^*$) are quite common and have been around (historically as "cholesterics") since the beginning of liquid crystal science. Moreover, several instances of the existence of chiral domains in nematics formed by a-chiral molecules have been reported.[3-8] On the other hand, the theoretical possibilities of a-chiral nematics with lower symmetry than the $N_u$ phase _ namely polar uniaxial nematics,[9-13] biaxial a-polar nematics[14-19] and polar biaxial nematics[12, 20-21] _ have been often proposed over the years. However, the experimental identification of such phases has thus far been rather rare[22-24] and in some cases controversial.[25-27]

The theoretical possibility of a biaxial nematic phase was demonstrated, first by M. J. Freiser, in 1970.[14] The prediction envisaged an a-chiral, a-polar phase with three mutually orthogonal symmetry axes. This corresponds to the $D_{2h}$ symmetry group and constitutes the most natural choice for illustrating the theoretical possibility, in the sense that it introduces the minimal asymmetry to the $N_u$ phase (i.e. the minimal symmetry breaking of $D_{\infty h}$) that could be reflected on the second rank tensor physical properties characterising an a-polar nematic phase. Understandably, most of the subsequent theoretical and computer simulation works on biaxial nematics[19] dealt, either explicitly or implicitly, with phases of this high symmetry. The same $D_{2h}$ symmetry was adopted for the consistent analysis of the experimental data on the first real biaxial nematic phase, discovered in 1980 by L. J. Yu and A. Saupe in a lyotropic system.[28] A $D_2$ symmetry, allowing for three mutually orthogonal two-fold symmetry axes and possible chiral asymmetry, was shortly thereafter adopted by A. Saupe[29] in his theoretical study of the elastic and flow properties of biaxial nematics. Also, the analysis of phase biaxiality observed in side-chain nematic liquid crystal polymers by optical measurements,[30] and later by NMR,[31] was based on the understanding that the investigated biaxial nematic phases have $D_{2h}$ symmetry and led to qualitatively consistent and quantitatively reasonable results. All this perhaps generated a tendency to design and interpret the majority of experiments in the long quest for the thermotropic biaxial nematic phase exclusively on the expectation that the sought phase is necessarily of orthorhombic symmetry, with the three symmetry axes defining a triplet of directors **n, l, m** that are common principal axes for all the macroscopic second-rank tensor properties of the medium. However, such restriction is neither warranted by theory, where a variety of symmetries for the biaxial nematic phase are *a priory* possible,[20, 32] nor imposed by experiment, particularly in the case of the recent compelling experimental evidence of biaxiality in bent-core[33] and calamitic-tetrapode[34, 35] nematics. The aim of the present work is to demonstrate that there are in fact strong experimental indications that phase biaxiality in these systems could be of lower symmetry than $D_{2h}$, to suggest ways of differentiating experimentally between biaxial nematic phases of different symmetries and to point out possible flaws in the interpretation of the experimental results in case a higher symmetry than the actual one is adopted for the analysis of the measurements on aligned samples.

The possible symmetries of an a-chiral, a-polar biaxial nematic phase are presented in section II, together with their general implications on the anisotropic physical properties of these materials. The consequences of the different symmetries on the alignment of biaxial nematics, particularly in relation to their electro-optic response are considered in section III. Deuterium NMR methods have thus far offered the most definitive means of identifying phase biaxiality in liquid crystals. A detailed analysis of these methods, focussing on the discriminating effects of different symmetries on the measurable spectra, is presented in section IV and in the two related appendices. The main results from section IV are carried over in section V to the application of infrared spectroscopy for the study of biaxial order in nematics. The presently available experimental results on the biaxiality of various nematic systems are critically reviewed in section VI in regards to their implications on the possible symmetries of the biaxial phase. The conclusions of the present study are collected in section VII.

## II. THE POSSIBLE SYMMETRIES OF BIAXIAL NEMATICS

For the purposes of the present work, a biaxial nematic liquid crystal is defined as a positionally disordered fluid that has at least one macroscopic second-rank tensor physical property $Q_{AB}$ exhibiting three different principal values $Q_{XX} \neq Q_{YY} \neq Q_{ZZ}$. For each such property, a set of unit vectors $\mathbf{n}^{(Q)}, \mathbf{l}^{(Q)}, \mathbf{m}^{(Q)}$, identified with the directions of the principal axes frame (*PAF*) of the tensor property is defined. In the absence of symmetry axes or planes, each independent tensor property of the medium has in principle its own distinct *PAF*. Such a medium is termed as polar biaxial if, in addition to the non-vanishing second-rank tensor properties, it has at least one non vanishing macroscopic vector property **p**. Clearly, the existence of an axis of higher than two-fold rotational symmetry renders the phase uniaxial and, in that case, only vector quantities in the direction of the symmetry axis can survive (polar uniaxial nematics). Conversely, a non-vanishing vector quantity whose direction does not coincide with that of a symmetry axis necessarily implies phase biaxiality together with phase polarity.

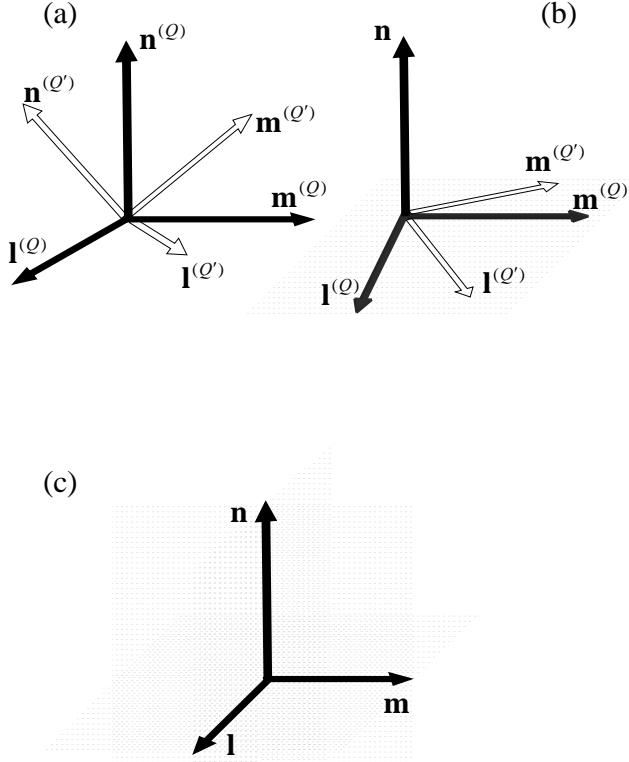

**FIG. 1**. Possible symmetries of the a-chiral, a-polar biaxial nematic phase: (a) The triclinic biaxial nematic phase, ($N_{Bt}$), corresponding to the $C_i$ point group; no unique director. (b) The monoclinic biaxial nematic phase, ($N_{Bm}$), corresponding to the $C_{2h}$ point group; one unique director (twofold symmetry axis). (c) The orthorhombic biaxial nematic, ($N_{Bo}$), corresponding to the $D_{2h}$ point group; a unique triplet of directors.

Restricting our considerations to a-chiral, a-polar biaxial nematics, the lowest possible symmetry for such phases corresponds to the triclinic point group $C_i$. In this case, the triclinic biaxial phase ($N_{Bt}$) has no unique director (Fig. 1(a)) and different macroscopic tensor properties $Q, Q'$ define in general different triplets of unit vectors $\left(\mathbf{n}^{(Q)}, \mathbf{l}^{(Q)}, \mathbf{m}^{(Q)}\right)$, $\left(\mathbf{n}^{(Q')}, \mathbf{l}^{(Q')}, \mathbf{m}^{(Q')}\right)$, with the inversion symmetry implying the "a-polarity" of any such triplet, i.e. the physical equivalence of $\left(\mathbf{n}^{(Q)}, \mathbf{l}^{(Q)}, \mathbf{m}^{(Q)}\right)$ to $\left(-\mathbf{n}^{(Q)}, -\mathbf{l}^{(Q)}, -\mathbf{m}^{(Q)}\right)$. The next level of symmetry for the a-polar, a-chiral biaxial nematics corresponds to the $C_{2h}$ point group. In this case, the monoclinic biaxial phase, ($N_{Bm}$), has one unique director that coincides with the twofold symmetry axis and is a common principal axis for all the second-rank tensor properties of the system (Fig. 1(b)). The other two principal axes for each such property are on the symmetry plane perpendicular to the twofold axis but are otherwise oriented differently for different tensor properties. The phase is separately symmetric with respect to the inversion of the unique director and the inversion of any direction on the symmetry plane. The only possibility of higher than $C_{2h}$ symmetry for the a-polar, a-chiral biaxial nematic phase is that of the orthorhombic $D_{2h}$ point group. It is only in this case that the phase has a unique triplet of directors $\mathbf{n}, \mathbf{l}, \mathbf{m}$ (coincident with the three twofold axes) forming the common *PAF* for all the second rank tensor macroscopic properties (Fig. 1(c)). Furthermore, all the physical properties of this orthorhombic biaxial phase, ($N_{Bo}$), are invariant with respect to the separate inversion of each one of the three directors $\mathbf{n}, \mathbf{l}, \mathbf{m}$. Lastly, on the grounds of 2$^{nd}$ rank tensor properties alone, a $D_{nh}$ symmetry with $n > 2$ qualifies the phase as uniaxial.

## III. THE ALIGNMENT OF BIAXIAL NEMATICS

Clearly, of the three biaxial a-polar, a-chiral types of nematics described in the previous section, namely the triclinic $N_{Bt}$, the monoclinic $N_{Bm}$ and the orthorhombic $N_{Bo}$, full three-dimensional alignment with respect to all the anisotropic properties of the medium along a common set of thee orthogonal directions is possible to achieve only in the latter. In other words, aligning any two of the directors of an $N_{Bo}$ medium (say aligning one director by surface anchoring and another by application of an external electric field in an orthogonal direction) automatically fixes uniquely the directions of the three principal axes for all the tensor properties of the medium. In the case of an $N_{Bt}$ medium, aligning two principal axes in mutually orthogonal directions will in general not be possible geometrically if these axes correspond to different tensors (say $\mathbf{n}^{(Q)}$ and $\mathbf{m}^{(Q')}$ of Fig. 1(a)). On the other hand, if the two axes belong to the same triplet (say $\mathbf{n}^{(Q)}$ and $\mathbf{m}^{(Q)}$ of Fig. 1(a)) and their simultaneous alignment in mutually orthogonal directions is achieved by some means, this will not imply that any of the axes belonging to a different triplet (say $\mathbf{n}^{(Q')}, \mathbf{l}^{(Q')}$ or $\mathbf{m}^{(Q')}$ in Fig. 1(a)) will be found parallel to either of the aligned directors or normal to the plane formed by them.

For an $N_{Bm}$ medium, the simultaneous orthogonal alignment of the single symmetry axis of the phase (say $\mathbf{n}$ in Fig. 1(b)) and any of "transverse" principal axes ($\mathbf{l}^{(Q)}, \mathbf{m}^{(Q)}, \mathbf{l}^{(Q')}, \mathbf{m}^{(Q')}$ in Fig. 1(b)) is obviously always possible geometrically but, unlike the $N_{Bo}$ medium, this does not single out on the symmetry plane two common directions for all the tensor properties. Thus, for example, the degenerate-planar anchoring of the symmetry axis $\mathbf{n}$ (Fig. 2(a)) on a substrate with simultaneous electric field alignment of the static dielectric principal axis $\mathbf{m}^{(es)}$ of the medium perpendicular to the substrate, would lead to a distribution of the transverse optical axes $\mathbf{m}^{(opt)}, \mathbf{l}^{(opt)}$ (essentially the principal axes of the optical frequency dielectric tensor, which need not coincide with the respective axes of the static dielectric tensor) on the surfaces of two cones about the symmetry axis, as shown in Fig. 2(a). Similarly, the non-degenerate homeotropic alignment of $\mathbf{n}$ perpendicular to a flat substrate carrying a "rubbing direction" $\mathbf{R}$ (see Fig. 2(b)) will not necessarily bring any of the transverse static dielectric axes, $\mathbf{m}^{(es)}, \mathbf{l}^{(es)}$, or of the optical ones $\mathbf{m}^{(opt)}, \mathbf{l}^{(opt)}$ in alignment with the rubbing direction.

The experimental results in all optic and electro-optic investigations[36-37] that, to our knowledge, have been carried out to date on the biaxiality of bent-core and tetrapode thermotropic nematics have been analyzed on the assumption of an orthorhombic symmetry, in which case the shift angles (such as $u^{(es,opt)}$ in Fig. 2) between the transverse principal axes of different tensor properties are forced to the "orthorhombic" values *0* or *π/2*. Thus, we are not aware of any experimental proof that that the *u*-angles in these systems are indeed *0* or *π/2* nor of any attempts to detect deviations of the *u*-angles from the orthorhombic values. Such deviations would clearly constitute direct proof of phase biaxiality, even if the magnitudes of the transverse optical or electric anisotropies were too small to measure unambiguously. Shift angles between principal axes are quite common in Smectic C *(SmC)* liquid crystals. These tilted and layered systems exhibit $C_{2h}$ phase biaxiality, with the twofold axis perpendicular to the tilt plane. Tilt-angle measurements are known to give appreciable, and temperature dependent, differences between the tilt-angle values obtained from optical, *X*-ray, *NMR* etc[38-44] methods on the same *SmC* compound, as well as differences between the mesogenic-core and pendant alkyl chain tilt-angles.[41-44] Furthermore, different tilt angles, and therefore distinct *PAF*s, for different properties of the same compound are obtained as direct theoretical results from molecular models in which internal flexibility together with the characteristic aliphatic-aromatic molecular composition of the *SmC* compounds are explicitly taken into account.[45-47] In the present context, these differences in the tilt angles obviously correspond to the shift-angles among the transverse principal axes associated with anisotropic properties of the $C_{2h}$ biaxial nematic medium.

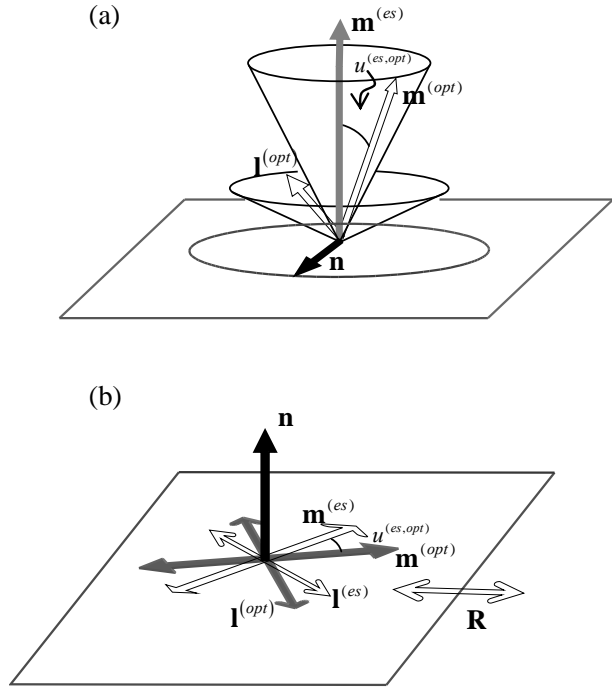

**FIG. 2.** Possible alignment of the monoclinic, $N_{Bm}$, biaxial nematic phase: (a) Degenerate-planar alignment of the symmetry axis **n** on a flat substrate simultaneously with the alignment of the dielectric principal axis $\mathbf{m}^{(es)}$ perpendicular to the substrate leads to the distribution of the optical axes $\mathbf{m}^{(opt)}$ and $\mathbf{l}^{(opt)}$ on the surfaces of two cones with apertures $u^{(es,opt)}$ and $\pi/2 - u^{(es,opt)}$, respectively. (b) Non-degenerate homeotropic alignment of **n** perpendicular to the rubbing direction **R** of the flat substrate does not necessarily bring $\mathbf{m}^{(opt)}$ or $\mathbf{l}^{(opt)}$ in alignment with **R**.

The above considerations underline the importance of proving, rather than simply assuming, the particular symmetry of the investigated phase biaxiality in optic or eletro-optic experiments on the new, low molar mass, thermotropic candidates of biaxial nematics. However, these types of experiments may not turn out to provide the most sensitive means for the distinction between the possible symmetry types of the phase biaxiality, either because of inherent methodological complexities related to surface alignment, director gradients etc, which could make a slightly deformed uniaxial system appear as biaxial,[48] or because the deviation angles among optical, dielectric and surface anchoring directions, might not be sufficiently large to allow a clear differentiation between the possible point group symmetries. On the other hand, the manifestations of these symmetries can be more pronounced on the orientational ordering of certain molecular segments, rather than on the global molecular ordering underlying the optical dielectric and anchoring anisotropies. It is then advantageous to use experimental methods, such as *NMR* or *IR* absorbance, which are suitable for the measurement of the orientational order of specifically labeled molecular sites. These are considered in the next two sections.

## IV. BIAXIAL NEMATIC PHASE SYMMETRIES AND $^2$*H-NMR* SPECTRA

*NMR* methods are widely used for the study of orientational order in liquid crystals[49-50] and are considered to offer a reliable way for the characterization of phase biaxiality. In particular, $^2$*H-NMR* methods, involving either directly deuteriated nematic molecules or deuteriated probe-molecules, have been used extensively for the study of biaxiality in nematics.[28, 31, 33, 35,]

[51-55] The measurable quantities in these methods are the frequency splittings $\delta v^{(i)}$ associated with the interaction between the deuterium quadrupole moment and the electric field gradient *(EFG)* at each of the deuterated sites *i* of the molecule in the presence of an applied magnetic field. The dependence of $\delta v^{(i)}$ on the time-averaged orientation of the deuterated molecular segment relative to the magnetic field is given by: [49, 56]

$$\delta v^{(i)} = \frac{3}{2} v_Q^{(i)} \left[ \left( \frac{3}{2} \left\langle \left( \hat{h} \cdot \hat{z}_i \right)^2 \right\rangle - \frac{1}{2} \right) + \frac{\eta_{EFG}^{(i)}}{2} \left( \left\langle \left( \hat{h} \cdot \hat{x}_i \right)^2 \right\rangle - \left\langle \left( \hat{h} \cdot \hat{y}_i \right)^2 \right\rangle \right) \right]. \quad (1)$$

Here, $v_Q^{(i)}$ is the quadrupolar coupling constant for the deuterated site *i* (its numerical value is usually obtained from measurements in the solid state), the angular brackets indicate time-averaging, the unit vector $\hat{h}$ denotes the direction of the magnetic field and the unit vectors $\hat{x}_i, \hat{y}_i, \hat{z}_i$ denote the principal axes of the *EFG* tensor $V_{ab}^{(i)}$ at the deuterated site *i*. The assignment of these axes is normally taken to correspond to ascending absolute magnitudes of the principal values, i.e. so that $|V_{x_i x_i}| \leq |V_{y_i y_i}| \leq |V_{z_i z_i}|$. Accordingly, the biaxiality of the *EFG* tensor, defined as

$$\eta_{EFG}^{(i)} \equiv \frac{V_{x_i x_i} - V_{y_i y_i}}{V_{z_i z_i}}, \quad (2)$$

is restricted in the range $0 \leq \eta_{EFG}^{(i)} \leq 1$. The actual values of $\eta_{EFG}^{(i)}$ are quite small and in most cases of practical interest the *EFG* is taken to have cylindrical symmetry around the deuterium bond direction,[57] implying $\eta_{EFG}^{(i)} = 0$.

To obtain the explicit dependence of $\delta v^{(i)}$ in Eq. (1) on the macroscopic orientation of the sample, a phase-fixed macroscopic frame *X,Y,Z*, is introduced and a second-rank symmetric and traceless tensor $G_{AB}^{(i)}$, describing the orientational averaging of the field gradient associated with the molecular site *i*, is defined with components

$$G_{AB}^{(i)} \equiv \frac{3}{2} \left\langle \left( \hat{A} \cdot \hat{z}_i \right) \left( \hat{B} \cdot \hat{z}_i \right) \right\rangle - \frac{1}{2} + \frac{\eta_{EFG}^{(i)}}{2} \left( \left\langle \left( \hat{A} \cdot \hat{x}_i \right) \left( \hat{B} \cdot \hat{x}_i \right) \right\rangle - \left\langle \left( \hat{A} \cdot \hat{y}_i \right) \left( \hat{B} \cdot \hat{y}_i \right) \right\rangle \right), \quad (3)$$

where the unit vectors $\hat{A}, \hat{B}$ represent the directions of the macroscopic axes *X,Y,Z*. Then, with the direction of the magnetic field in the macroscopic frame given by the unit vector components $h_A$, the splitting can be written as

$$\delta v^{(i)} = \frac{3}{2} v_Q^{(i)} h_A h_B G_{AB}^{(i)}, \quad (4)$$

with summation implied over the repeated tensorial indices *A,B*.

The choice of the macroscopic frame is obvious in the case of the $N_{Bo}$ phase, where the presence of three orthogonal symmetry axes defines uniquely the principal axes of all second rank tensors, and therefore of $G_{AB}^{(i)}$ for any site *(i)*. However, for an $N_{Bm}$ or $N_{Bt}$ phase, careful distinction should be made between two macroscopic frames, both of which are directly relevant to the complete analysis of the *NMR* measurements. The one frame is specific to quadrupolar interaction associated with the deuterated site and is defined as the *PAF* of the $G_{AB}^{(i)}$ tensor (the $G^{(i)} - PAF$, for brevity) and the other is the principal axis frame of the diamagnetic susceptibility tensor of the phase (the $\chi^m - PAF$). The relevance of these two frames is described in some detail below.

## A. The quadrupolar interaction frame

In the $G^{(i)} - PAF$, with its axes denoted by $X_i, Y_i, Z_i$, the tensor $G^{(i)}_{AB}$ is diagonal. The assignment of the principal axes is chosen according to ascending order of the absolute magnitudes of the principal values, $\left|G^{(i)}_{X_iX_i}\right| \leq \left|G^{(i)}_{Y_iY_i}\right| \leq \left|G^{(i)}_{Z_iZ_i}\right|$. Therefore, $G^{(i)}_{AB}$ in this frame can be fully described in terms of the primary order component

$$S^{(i)} \equiv G^{(i)}_{Z_iZ_i} \tag{5}$$

and the biaxiality parameter

$$\eta^{(i)} \equiv \frac{G^{(i)}_{X_iX_i} - G^{(i)}_{Y_iY_i}}{G^{(i)}_{Z_iZ_i}} \quad . \tag{6}$$

The primary component $S^{(i)}$ becomes equal to 1 at perfect segmental order (coincidence of the molecular-segment axes $\hat{x}_i, \hat{y}_i, \hat{z}_i$ with the macroscopic axes $X_i, Y_i, Z_i$). The parameter $\eta^{(i)}$ quantifies the biaxiality of the phase as reflected on the orientational order of the deuterated site $i$ and, according to the assignment of the principal axes, its values are restricted in the range $0 \leq \eta^{(i)} \leq 1$. Under the usual assumption of cylindrically symmetric *EFG* around the deuterium bond ($\eta^{(i)}_{EFG} = 0$), the primary component and the biaxiality parameter can be expressed according to Eqs. (3), (5) and (6) in terms of the azimuthal ($\alpha_i$) and polar ($\beta_i$) angles of the bond direction $z_i$ relative to the $G^{(i)} - PAF$ as

$$S^{(i)} = \left\langle \frac{3}{2}\cos^2\beta_i - \frac{1}{2} \right\rangle \quad ; \quad \eta^{(i)} = \frac{\left\langle \sin^2\beta_i \cos 2\alpha_i \right\rangle}{\frac{2}{3} - \left\langle \sin^2\beta_i \right\rangle} \quad . \tag{7}$$

It then strictly follows from these relations that the physically acceptable values of $S^{(i)}$ and $\eta^{(i)}$ are constrained by the following condition:

$$S^{(i)}\left(\eta^{(i)} + 1\right) \leq 1 \quad . \tag{8}$$

Irrespectively of any assumptions on the smallness of $\eta^{(i)}_{EFG}$, the values of $S^{(i)}$ and $\eta^{(i)}$ can be determined from the values of the splittings at two independent orientations of the magnetic filed relative to the $G^{(i)} - PAF$. With the angles $\theta_i, \phi_i$ describing the orientation of $\hat{h}$ in the $X_i, Y_i, Z_i$ axis system, Eq. (4) assumes the form

$$\delta\nu^{(i)} = \frac{3}{2}\nu_Q^{(i)} S^{(i)} \left[ \left(\frac{3}{2}\cos^2\theta_i - \frac{1}{2}\right) + \frac{\eta^{(i)}}{2}\sin^2\theta_i \cos 2\phi_i \right] \quad . \tag{9}$$

The dependence of $\delta\nu^{(i)}$ on the angles $\theta_i, \phi_i$ presents two extrema. The one, $\delta\nu_1^{(i)} = \frac{3}{2}\nu_Q^{(i)} S^{(i)}$, is obtained when the magnetic field is directed along the $Z_i$ axis (i.e. for $\cos^2\theta_i = 1$) and the other, $\delta\nu_2^{(i)} = -\frac{3}{4}\nu_Q^{(i)} S^{(i)}\left(1 + \eta^{(i)}\right)$, is obtained when the magnetic field is directed along $Y_i$ (i.e. for $\sin^2\theta_i \cos 2\phi_i = -1$). Given the numerical value of the quadrupolar coupling constant $\nu_Q^{(i)}$, the first exteremum yields directly the primary component

$$S^{(i)} = 2\delta\nu_1^{(i)} / 3\nu_Q^{(i)} \quad , \tag{10}$$

while the biaxiality parameter is obtained from the ratio of two extrema,

$$\eta^{(i)} = -\left(1 + 2\frac{\delta v_2^{(i)}}{\delta v_1^{(i)}}\right) \quad . \tag{11}$$

Clearly, $\delta v_1^{(i)} \leq -2\delta v_2^{(i)}$, with the equality holding for all the sites $i$ in the case of a uniaxial phase.

Whilst the $G^{(i)} - PAF$ of a deuterated site $i$ is the natural choice for the description of the orientational order of that site, the orientation of the liquid crystal sample relative to the magnetic filed in the actual *NMR* experiment is normally determined by the spectrometer magnetic field itself which, due to the considerable diamagnetic anisotropy of most liquid crystals, becomes the primary aligning stimulus in high-field *NMR* experiments. This makes it necessary to introduce a macroscopic frame of axes, the $\chi^m - PAF$, that is suitable for the description the magnetic alignment of the sample.

## B. The magnetic susceptibility frame

In the $\chi^m - PAF$, with its phase-fixed axes denoted by $X_M, Y_M, Z_M$, it is the magnetic susceptibility tensor $\chi^m_{AB}$ that becomes diagonal. Assigning the axes according to the ascending sequence $\chi^m_{X_M X_M} \leq \chi^m_{Y_M Y_M} \leq \chi^m_{Z_M Z_M}$ of the principal values, the minimum magnetic energy of an unconstrained monodomain sample will be obtained when the magnetic field is along the $Z_M$ axis. If, on the other hand, this axis is by some means made to form an angle $\theta_M$ with the magnetic field, then the energy will be minimized for $X_M$ directed perpendicular to the plane formed by the direction of the magnetic field and the $Z_M$ axis, i.e. for the values $\phi_M = \pm \pi/2$ of the azimuthal angle of the magnetic field in the $\chi^m - PAF$.

For the $N_{Bt}$ phase, the $\chi^m - PAF$ has in general no common axis with the $G^{(i)}$-PAF and therefore the right hand side of Eq. (4) for the splitting, expressed in the $\chi^m - PAF$, will involve the five independent components of the $G^{(i)}_{AB}$ tensor in that frame, namely

$$\delta v^{(i)}(\theta_M, \phi_M) = \frac{3}{2} v_Q^{(i)} \left[ \begin{array}{l} \left(\frac{3}{2}\cos^2\theta_M - \frac{1}{2}\right) G^{(i)}_{Z_M Z_M} + \frac{1}{2}\sin^2\theta_M \cos 2\phi_M \left(G^{(i)}_{X_M X_M} - G^{(i)}_{Y_M Y_M}\right) \\ + \sin 2\theta_M \left(\cos\phi_M G^{(i)}_{Z_M X_M} + \sin\phi_M G^{(i)}_{Z_M Y_M}\right) + \sin^2\theta_M \sin 2\phi_M G^{(i)}_{Y_M X_M} \end{array} \right] \tag{12}$$

First we note that, according to this equation, for the magnetically aligned sample, i.e. with the $Z_M$ parallel to the magnetic field, the splitting has the value,

$$\delta v_\parallel^{(i)} = \frac{3}{2} v_Q^{(i)} G^{(i)}_{Z_M Z_M} \quad . \tag{13}$$

Similarly, the value of the splitting when the magnetic field is directed along the $Y_M$ axis is

$$\delta v_\perp^{(i)} = -\frac{3}{4} v_Q^{(i)} \left[ G^{(i)}_{Z_M Z_M} + \left(G^{(i)}_{X_M X_M} - G^{(i)}_{Y_M Y_M}\right)\right]. \tag{14}$$

Thus, if both the splittings $\delta v_\parallel^{(i)}$ and $\delta v_\perp^{(i)}$ can be measured, the values of the parameters

$$S_M^{(i)} \equiv G^{(i)}_{Z_M Z_M} = 2\delta v_\parallel^{(i)} / 3 v_Q^{(i)} \tag{15}$$

and

$$\eta_M^{(i)} \equiv \left( G_{X_M X_M}^{(i)} - G_{Y_M Y_M}^{(i)} \right) / G_{Z_M Z_M}^{(i)} = -\left(1 + 2\left(\delta v_\perp^{(i)} / \delta v_\parallel^{(i)}\right)\right) \qquad (16)$$

will be obtained. These two parameters have an obvious formal correspondence to the primary order parameter $S^{(i)}$ and the biaxiality parameter $\eta^{(i)}$ appearing in Eqs. (10) and (11). However, unless the axes $X_M, Y_M, Z_M$ happen to coincide respectively with $X_i, Y_i, Z_i$ — a coincidence for which there is no physical reason in a biaxial phase of monoclinic or triclinic symmetry — the frequencies $\delta v_\parallel^{(i)}$, $\delta v_\perp^{(i)}$ are not equal to the frequency extrema $\delta v_1^{(i)}$, $\delta v_2^{(i)}$ in Eqs. (10) and (11), and, as shown below, the values of $S_M^{(i)}$ and $\eta_M^{(i)}$ could vastly differ from the values of the order parameter $S^{(i)}$ and the biaxiality parameter $\eta^{(i)}$, respectively. In what follows, $S_M^{(i)}$ and $\eta_M^{(i)}$ will be referred to as the "apparent" parameters, to stress their distinction from the true parameters $S^{(i)}$ and $\eta^{(i)}$.

According to Eqs. (15) and (16), the apparent parameters $S_M^{(i)}$ and $\eta_M^{(i)}$ are the directly measurable quantities in those experiments where separate measurements of the splittings parallel and perpendicular to the magnetic field are possible. This is usually the case for systems with sufficiently long relaxation times of the magnetic reorientation.[28, 31, 35, 51] However, knowledge of the apparent parameters $S_M^{(i)}$ and $\eta_M^{(i)}$ is not enough for the evaluation of the true parameters $S^{(i)}$ and $\eta^{(i)}$ except when the biaxial phase is known to be of $N_{Bo}$ symmetry, in which case the off-diagonal components of $G_{AB}^{(i)}$ vanish in the $\chi^m - PAF$ as well. Furthermore, in those cases where a $\theta_M = \pi/2$ configuration of the sample can be achieved, the angle $\phi_M$ is often not sharply restricted to the energy minimising values $\pm \pi/2$ but, due to relatively weak magnetic biaxiality, may exhibit a distribution over its full range.[27, 31, 35] This gives rise to a distribution of splittings in the range $-\frac{3}{4} v_Q^{(i)} S_M^{(i)} \left(1 \pm \eta_M^{(i)}\right)$, rather than the single splitting of Eq. (14). In this case, the relative width of the $\delta v_\perp^{(i)}$ distribution provides a measure of the apparent biaxiality parameter $\eta_M^{(i)}$.

## C. Differentiating between point group symmetries

The general relations between the apparent and the proper parameters are summarized in Appendix I. In the case of $N_{Bo}$ phase symmetry, these relations provide six alternative assignments of $S^{(i)}$ and $\eta^{(i)}$ for a given pair $S_M^{(i)}$ and $\eta_M^{(i)}$ (see Eq. (AI.3)). These alternatives correspond to the six possibilities of matching the $X_i, Y_i, Z_i$ with the $X_M, Y_M, Z_M$ axes. The acceptable assignment is to be chosen subject to restriction imposed by the definitions of $S^{(i)}$ and $\eta^{(i)}$, in particular $S^{(i)} \leq 1$ and $0 \leq \eta^{(i)} \leq 1$, and the mutual compatibility constraints of the type implied by Eq. (8). The six matching alternatives between the $\chi^m - PAF$ and the $G^{(i)}$-$PAF$ axes of the $N_{Bo}$ are the direct generalization of the possibility of having positive or negative diamagnetic anisotropy in the uniaxial nematics. There, the $Z_i$ axis necessarily coincides with the director **n** while the axis of magnetic alignment can be either along **n** (i.e positive diamagnetic anisotropy, resulting in $Z_i \parallel Z_M$ and therefore $S_M^{(i)} = S^{(i)}$; $\eta_M^{(i)} = 0$) or perpendicular to **n** (i.e negative diamagnetic anisotropy, resulting in $Z_i \perp Z_M$ and $S_M^{(i)} = -S^{(i)}/2; \eta_M^{(i)} = \pm 3$). A crossover from $Z_i \parallel Z_M$ to $Z_i \perp Z_M$ by varying the temperature within the same biaxial nematic phase has been reported for the lyotropic system in Ref. 51

For $N_{Bm}$ or $N_{Bt}$ phase symmetry, the determination of $S^{(i)}$ and $\eta^{(i)}$ requires, according to Eqs. (AI.1) and (AI.2) the knowledge of the off-diagonal components of $G_{A_M B_M}^{(i)}$, in addition to $S_M^{(i)}$ and $\eta_M^{(i)}$. In the cases where the splitting can be measured for a sufficiently large set of different orientations $\theta_M, \phi_M$ (either individually or in a well defined superposition, such as a 3-dimmensional "powder"), the five tensor components appearing in the expression of Eq. (12) can be evaluated and from those one can obtain the order parameter $S^{(i)}$ and the biaxiality $\eta^{(i)}$ for the site $i$ as well as the orientations of the respective principal axes $X_i, Y_i, Z_i$ relative to the axes $X_M, Y_M, Z_M$. If such complete angular dependence data for the splitting cannot be obtained, as it is normally the case with low-molar-mass and/or low viscosity liquid crystals, the tensor components $G_{Z_M Z_M}^{(i)}, \left(G_{X_M X_M}^{(i)} - G_{Y_M Y_M}^{(i)}\right)$ and $G_{Z_M Y_M}^{(i)}$ can be obtained by a method combining measurements of the splittings of the statically aligned sample in the magnetic field with measurements of the spectral pattern generated by spinning the sample about an axis perpendicular to the magnetic field. This method has been used for the identification of biaxial order in several instances of $SmC$[58-61] and nematic[33,53] phases where it is difficult to achieve a predetermined static orientation of the sample away from its minimum-energy direction in the magnetic field. The formulation of the method is outlined in Appendix II. From the phase-symmetry viewpoint, the following points should be noted regarding its application to biaxial nematics:

(i) As a result of the alignment of the $X_M$ axis along the spinning axis ($\cos\phi_M = 0$ in Eq. (12)), the spectral pattern has no information on the components $G_{Z_M X_M}^{(i)}, G_{Y_M X_M}^{(i)}$. Accordingly, the combined static/spinning sample measurements can lead to a complete determination of the $G_{A_M B_M}^{(i)}$ tensor only if $X_M$ (i.e the axis of lowest diamagnetic susceptibility) happens to be a symmetry axis of the phase, in which case both $G_{Z_M X_M}^{(i)}$ and $G_{Y_M X_M}^{(i)}$ would vanish by symmetry. In that case, the measurable angle $g^{(i)}$ in Eqs. (AII.1, AII.2) would simply be the angle by which the $\chi^m - PAF$ has to be rotated about the $X_M$ axis in order for its other two axes, $Y_M, Z_M$, to be brought in coincidence with two of the $X_i, Y_i, Z_i$ axes of the $G^{(i)}$-PAF (the third one of these axes necessarily coincides with the symmetry axis $X_M$; see Table I in Appendix I). Regarding the site-dependence and temperature-dependence of the $g^{(i)}$ angles, it should be noted that the $G^{(i)}$-PAFs of inequivalent sites will, in principle, be oriented differently relative to one-another and to the $\chi^m - PAF$; moreover, for flexible molecules, the orientations of the different $G^{(i)}$-PAFs with respect to the $\chi^m - PAF$ could vary differently with temperature as a result of differences in the conformational motions of the various molecular segments. Lastly, if $X_M$ it is not known to be a symmetry axis, the rotation about $X_M$ alone would clearly not be enough to bring the $G^{(i)}$-PAF in coincidence with the $\chi^m - PAF$ since additional rotations would be required, associated with the non vanishing, albeit inaccessible to this measurement, components $G_{Z_M X_M}^{(i)}, G_{Y_M X_M}^{(i)}$.

(ii) The component $G^{(i)}_{Z_M Y_M}$ would vanish if one of the $Y_M$, $Z_M$, axes happened to be a symmetry axis of the phase. Therefore, the measurement would yield $\sin 2g^{(i)} = 0$ (see Eq. AII.2). Clearly, however, this alone, would not imply that the $\chi^m - PAF$ shares the same axes with the $G^{(i)}$-PAF. If, for example, the phase has a single symmetry axis coinciding with the direction of $Z_M$, then the $G^{(i)}_{Y_M X_M}$ component would in general be non-zero and this would imply that the $\chi^m - PAF$ and the $G^{(i)}$-PAF are rotated relative to one another about their common axis $Z_M$ by an angle $g^{(i)}_{Z_M}$ that cannot be evaluated from the combined static/spinning sample measurements. Consequently, a measurement in which the result $G^{(i)}_{Z_M Y_M} = 0$ is obtained would specify the $G^{(i)}_{A_M B_M}$ completely only if it is known that both $Y_M$ and $Z_M$ are symmetry axes of the phase (i.e. if the phase is $N_{Bo}$).

(iii) Whatever the symmetry of the biaxial phase, the combined static/spinning sample measurements can always yield the values of the apparent parameters $S^{(i)}_M$ and $\eta^{(i)}_M$. However, the relation of these parameters to the true, $S^{(i)}$ and $\eta^{(i)}$, depends not only on the symmetry of the phase but also on the specific associations of the $X_M, Y_M, Z_M$ and $X_i, Y_i, Z_i$ with the symmetry axes of the phase (when present) and with one another. These relations are given in Eqs. (AI.3) for the case of an orthorhombic biaxial phase ($D_{2h}$) and are summarized in Table I of Appendix I for a biaxial phase of monoclinic symmetry ($C_{2h}$). The respective detailed relations in the case of triclinic symmetry cannot be established within the combined static/spinning sample measurement scheme due to its aforementioned inability to provide the values of the components $G^{(i)}_{Z_M X_M}, G^{(i)}_{Y_M X_M}$. Accordingly, for triclinic phase symmetry, or monoclinic in which the maximum magnetic energy axis $X_M$ does not coincide with the symmetry axis of the phase, a technique allowing the recording of $\delta v^{(i)}$ for configurations in which neither $X_M$ is perpendicular nor $Z_M$ is parallel to the magnetic field would be required for the determination of $S^{(i)}$ and $\eta^{(i)}$. For example, in those cases where a sample with $\theta_M = \pi/2$ and $\phi_M$ distributed can be achieved, the spectral pattern yields, in addition to $S^{(i)}_M$ and $\eta^{(i)}_M$, the value of the $G^{(i)}_{Y_M X_M}$ component.

To assess the implications of a monoclinic symmetry on the apparent parameters $S^{(i)}_M$ and $\eta^{(i)}_M$ and their relation to the proper ones $S^{(i)}$ and $\eta^{(i)}$, we consider in some detail one of the of different symmetry axis assignments listed in Table I of Appendix I. Analogous implications apply for the other assignments. We choose the case where the principal axes $X_M$ and $X_i$ coincide with the twofold axis (top left cell of Table I) of the phase for which the angle $u^{(Z_M, Z_i)} = g^{(i)}$ together with $S^{(i)}_M$ and $\eta^{(i)}_M$ can be measured with the static/spinning sample scheme. The important implications stem from the appearance of $u^{(Z_M, Z_i)}$ in the relations between apparent and proper parameters. Thus, according to the relation (second equation in the top left cell of Table I)

$$\eta^{(i)}_M = \frac{-3\left[(1-\eta^{(i)})/(3+\eta^{(i)})\right] + \cos 2u^{(Z_M, Z_i)}}{\left[(1-\eta^{(i)})/(3+\eta^{(i)})\right] + \cos 2u^{(Z_M, Z_i)}} \tag{17}$$

a large apparent biaxiality $\eta_M^{(i)}$ can be obtained even if the proper biaxiality $\eta^{(i)}$ is negligible. For example, for $u^{(Z_M,Z_i)} = 30°$ and $\eta^{(i)} = 0$, this equation yields an enormous apparent biaxiality $\eta_M^{(i)} = -0.6$ (note that, unlike $\eta^{(i)}$, the values of the apparent biaxiality $\eta_M^{(i)}$ are not restricted in the range 0 to 1). Conversely, a negligible apparent biaxiality may be measured even if the proper biaxiality is large in case the value of $\cos 2u^{(Z_M,Z_i)}$ happens to be close to $3(1-\eta^{(i)})/(3+\eta^{(i)})$. For example, a sizeable proper biaxiality of $\eta^{(i)} = 0.2$ would lead to negligible apparent biaxiality, $\eta_M^{(i)} \approx 0$, if the angle $u^{(Z_M,Z_i)}$ were around $20°$. Furthermore, $\eta_M^{(i)}$ may exhibit a markedly different temperature dependence from the one typically expected for $\eta^{(i)}$ due to the additional variation caused by the temperature dependence of the angle $u^{(Z_M,Z_i)}$, whose absolute value is normally expected to increase with decreasing temperature (increasing biaxial order of the phase). Perhaps more importantly, the apparent primary order parameter $S_M^{(i)}$ may exhibit "anomalous" temperature dependence, namely decrease with decreasing temperature, if the rate of increase of the magnitude of $u^{(Z_M,Z_i)}$ with decreasing temperature overcompensates for the rate of increase of the proper primary order parameter $S^{(i)}$. This can be readily demonstrated by differentiating the relation (first equation in top left cell in Table I of Appendix I)

$$S_M^{(i)} = S^{(i)}\left[\left(\frac{3}{2}\cos^2 u^{(Z_M,Z_i)} - \frac{1}{2}\right) - \frac{\eta}{2}\sin^2 u^{(Z_M,Z_i)}\right] \qquad (18)$$

with respect to the temperature to obtain

$$\frac{dS_M^{(i)}}{dT} = \left(\frac{dS^{(i)}}{dT}\right)\frac{S_M^{(i)}}{S^{(i)}} - \left(\frac{d\eta^{(i)}}{dT}\right)\frac{S^{(i)}\sin^2 u^{(Z_M,Z_i)}}{2} - \left(\frac{d(\sin^2 u^{(Z_M,Z_i)})}{dT}\right)\frac{S^{(i)}(3+\eta^{(i)})}{2} \quad . \quad (19)$$

Normally, the temperature derivatives of $S^{(i)}$ and of $\sin^2 u^{(Z_M,Z_i)}$ are negative while both signs are possible for the derivative of $\eta^{(i)}$. Accordingly, depending on the relative magnitudes of these derivatives and of their coefficients for each molecular site $i$, a positive (anomalous) temperature derivative of $S_M^{(i)}$ may be obtained for some sites and negative (regular) for others on the same molecule. Lastly, it follows directly from Eq. (18) that a negligible apparent primary order $S_M^{(i)}$ can be obtained for a site that has high $S^{(i)}$ if the angle $u^{(Z_M,Z_i)}$ for that site is close to satisfying $\sin^2 u^{(Z_M,Z_i)} = 2/(3+\eta^{(i)})$. In this case the apparent biaxiality $\eta_M^{(i)}$ acquires divergently large magnitudes. Such instances have been observed in the biaxial order of the *SmC* phase of conventional calamitic compounds,[60] where $u^{(Z_M,Z_i)}$ values as large $50°$ have also been measured.

The above symmetry considerations for quadrupolar splittings apply to dipolar coupling interactions as well as chemical shift asymmetry interactions: In phases of lower than $N_{Bo}$ symmetry, the *PAFs* associated with these interactions need not coincide with each other nor with the respective *PAFs* of the quadrupolar interactions or the $\chi^m$ – *PAF* of the phase. Conversely, a relative deviation among any of these *PAFs* constitutes direct proof of phase biaxiality. It should be noted, however, that appreciable relative deviations are to be expected between the *PAFs* of molecular segments that undergo substantially different motional averaging. As it is often the case in conventional calamitic *SmC*s, deuterated sites that follow closely the orientational ordering of the mesogenic core, which essentially determines the diamagnetic susceptibility tensor of the phase, show marginal deviations of their $G^{(i)}$-*PAF* from the $\chi^m - PAF$ of the phase, even for strongly tilted, and therefore biaxial, compounds.[60] The analysis of the spectral patterns in the presence of chemical shift asymmetry is outlined in Appendix II.

## V. MOLECULAR ORIENTATIONAL ORDER MEASUREMENTS BY POLARIZED *IR* SPECTROSCOPY

Polarized infrared (*IR*) absorption measurements have been used for the study of molecular-segment orientational order in connection with the identification of nematic phase biaxiality.[34] The absorption of the *IR* beam depends on the direction of its polarization relative to the orientation $\hat{\mu}^{(i)}$ of the transition dipole moment associated with the respective absorption band. The anisotropic part of the *IR* absorption in a nematic liquid crystal is conveyed by the second rank symmetric and traceless absorbance tensor $A_{AB}^{(i)}$, defined in terms of the orientational averages of the transition dipole moment, with components:

$$A_{AB}^{(i)} = A_0^{(i)} \frac{3}{2} \left\langle (\hat{\mu}^{(i)} \cdot \hat{A})(\hat{\mu}^{(i)} \cdot \hat{B}) \right\rangle - \frac{1}{2} \qquad . \qquad (20)$$

Here the superscript *(i)* refers to the molecular segment in which the transition dipole is situated, $A_0^{(i)}$ is the absorbance strength parameter and $\hat{A}, \hat{B}$ denote unit vectors along the axes *X,Y,Z* of a macroscopic phase-fixed frame. The absorbance tensor components are the fundamental measurable quantities in *IR* absorption experiments on oriented samples. The phase symmetry implications on the tensor $A_{AB}^{(i)}$ are analogous to those discussed in detail for the field-gradient tensor $G_{AB}^{(i)}$ underlying the $^2$*H-NMR* spectra, and may be summarised as follows:

A macroscopic principal axis frame (the $A^{(i)} - PAF$) can be defined for the absorbance tensor for each transition dipole moments $\mu^{(i)}$ on the molecular structure. The principal values of each absorbance tensor can be expressed in terms of a primary order parameter $S^{(i)} (\equiv A_{Z_i Z_i}^{(i)} / A_0^{(i)})$ and a biaxiality $\eta^{(i)} (\equiv (A_{X_i X_i}^{(i)} - A_{Y_i Y_i}^{(i)}) / A_{Z_i Z_i}^{(i)})$, whose values, with proper assignment of the principal axes, satisfy the constraints of Eq. (8). The $A^{(i)} - PAFs$ of different transition dipole moments $\mu^{(i)}$ will strictly coincide only if the biaxial phase is of the $N_{Bo}$ symmetry. For $N_{Bm}$ symmetry, the different $A^{(i)} - PAFs$ will have one common axis, the symmetry axis of the phase, and in general will differ by a rotation about that axis. The angle of this rotation can be determined experimentally by measuring the angular dependence of the absorbance of different bands about the symmetry axis. For $N_{Bt}$ symmetry, the $A^{(i)} - PAFs$ of distinct sites will in general have no common axis, and experimental determination of their relative orientations requires a full *3-D* angular dependence study of the absorbances. The absorbance measurements are obtained for different orientations relative to a macroscopic frame that is singled out by the sample alignment mechanism (electric, magnetic, surface anchoring etc). To maintain a close analogy with the preceding description of the *NMR* measurements, this frame will be denoted as the *M*-frame, with axes $X_M, Y_M, Z_M$. Except for the case of $N_{Bo}$ phase symmetry, this frame will in general differ from the $A^{(i)} - PAF$ of the measured band and therefore the measurement in the two orthogonal directions will yield the values of the "apparent" primary order and biaxiality parameters for that band, which may differ significantly from the values of the actual $S^{(i)}$ and $\eta^{(i)}$. The determination of the latter requires, in addition to the apparent values, knowledge of the orientation of the $A^{(i)}-PAF$ relative to the *M*-frame.

Rather than describing the orientational order of the molecules merely in terms of a collection of absorbance $A_{AB}^{(i)}$ tensors associated with different molecular sites, it is possible, under certain conditions, to construct the ordering tensor of the entire molecule by combining the information from the different segments. To this end, a common molecular frame *x,y,z*, and a common macroscopic frame *X,Y,Z* are singled out for all the molecular sites, and the ordering tensor $S_{AB}^{ab}$ of the molecule is formed from the average orientations of the molecular axes (collectively represented by the indices *a, b*) relative to the macroscopic frame axes (*A, B* indices) as $S_{AB}^{ab} = \frac{3}{2}\langle(\hat{a}\cdot\hat{A})(\hat{b}\cdot\hat{B})\rangle - \frac{1}{2}$. Then, under the assumption that the reorientation of the transition dipoles relative to the molecular frame are to a good approximation statistically independent of the reorientations of the molecular frame relative to the macroscopic frame, in short the "Reorientation Decoupling Approximation" (*RDA*), the absorbance tensors $A_{AB}^{(i)}$ for such transition dipoles can be expressed as:

$$A_{AB}^{(i)} = A_{ab}^{(i)} S_{AB}^{ab} \qquad (21)$$

In this expression $A_{ab}^{(i)} = A_0^{(i)} \frac{3}{2}\langle(\hat{\mu}^{(i)}\cdot\hat{a})(\hat{\mu}^{(i)}\cdot\hat{b})\rangle - \frac{1}{2}$ are the conformational components of the absorbance tensor, which are determined by the motional averaging of the transition dipole direction relative to the axes of the molecular frame. Knowledge of these tensors for a sufficiently number of independent transition dipoles would allow the determination of the molecular order tensor $S_{AB}^{ab}$ in terms of the measurable absorbance components $A_{AB}^{(i)}$. A quite analogous procedure can be applied for the field gradient tensor components $G_{AB}^{(i)}$ which are measurable in $^2$H-NMR experiments. The *RDA* is obviously valid for those of the transition dipoles that are rigidly fixed with respect to the molecular frame. However, in the presence of extensive conformational motions quantitative validity of the *RDA* becomes questionable and the simple approach based on Eq. (21) has to be replaced by a more accurate one, taking into account the correlations between conformational motions and molecular reorientations.[49]

Within the *RDA*, the number of independent components of the order tensor $S_{AB}^{ab}$ required in Eq. (21) for the complete description of the absorbances, depends on the symmetry of the phase and of the molecules. In the most symmetric case, wherein the molecules are cylindrically symmetric and the phase is $N_{Bo}$, the independent components are just two, which with the usual assignment of axes are $S \equiv S_{ZZ}^{zz}$ and $P \equiv S_{XX}^{zz} - S_{YY}^{zz}$. This case however, implying higher molecular symmetry than the symmetry of the phase would be applicable to solute molecules in a biaxial phase rather than the molecules that actually self organize to form such phase. The molecules that do form a biaxial nematic phase need to be of at least orthorhombic symmetry. In that case, two more independent components appear, conventionally chosen to be $D \equiv S_{ZZ}^{xx} - S_{ZZ}^{yy}$ and $C \equiv S_{XX}^{xx} - S_{XX}^{yy} - S_{YY}^{xx} + S_{YY}^{yy}$. For molecules of monoclinic symmetry forming a $N_{Bm}$ the number of independent components is 9 and in the most asymmetric combination of triclinic (a-chiral) molecules in a $N_{Bt}$ phase, the number of independent components of $S_{AB}^{ab}$ mounts to 25. Consequently, the practical applicability of the representation of the absorbances in terms of molecular order tensor $S_{AB}^{ab}$ is essentially restricted to high symmetry molecules in the $N_{Bo}$, even when the *RDA* is expected to hold reasonably well.

## VI. DISCUSSION

The presently available experimental results on biaxial nematics, viewed in the light of the theoretical symmetry considerations of sections II-V, lead us to the following inferences: The detailed *NMR* investigations on lyotropic biaxial nematics[28, 51] seem to consistently establish a $D_{2h}$ assignment for the symmetry of these systems. Also, the analysis of the existing optical[30] and NMR data[31] on the side-chain biaxial nematic polymers is not inconsistent with that symmetry. In contrast, an analogously consistent assignment has not been presently reached for the bent-core[33] or for the tetrapode thermotropic systems.[34] Specifically, the *$^2$H-NMR* measurements by the static/spinning sample technique on bent-core nematics[33, 52-53] have yielded marginal values for the biaxiality parameter $\eta$ of certain dueteriated sites on the molecular core. These particular sites, however, turned out to be insensitive to the biaxial order even in the *SmC* of the same compounds. On the other hand, measurements using deuteriated probe solutes have yielded quite reasonable $\eta$ values, on the order of 0.1, which signal the presence of biaxial order in the nematic phase of these compounds. Clearly, however, this set of observations is too limited to prove or disprove any of the possible symmetries for the biaxial nematic phase. According to the considerations in section IV, a definitive symmetry assignment on the basis of a *NMR* study would necessitate a more comprehensive set of measurement, combining quadrupolar splittings from several structurally different deuteriated sites together with dipolar coupling and possibly chemical shift asymmetry data.

The calamitic tetrapode nematics have been investigated for biaxial order by *$^2$H-NMR* and by *IR* spectroscopy. Deuteriated probe solutes where used for the NMR studies, which included measurements of splittings parallel and perpendicular to the aligning magnetic field[35] as well as measurements of spinning sample spectral patterns.[54] These studies have yielded unusually high values for the biaxiality parameter $\eta$ which are not physically compatible with the simultaneously measured values of the order parameter $S$ at the low temperature end of the biaxial nematic phase as they grossly violate the constraint imposed by Eq. (8) on the acceptable values of these parameters. Since, however, these values are obtained by means of an analysis which, in addition to adopting a $D_{2h}$ symmetry for the investigated phase, involves a sequence of other assumptions regarding the statistics and dynamics of motional averaging, it is difficult to trace back the source of incompatibility between the final results on $S$ and $\eta$. The same tetrapode compounds have also been studied by IR absorbance.[34] The measurements where analyzed on the basis of orthorhombic symmetry both, for the phase an for the molecular structure, and, following the method of Eq. (21), the results where expressed in terms of the four molecular order parameters identified in section V for that combination of symmetries. The calculated values of the order parameters $S$ and $P$ appear to be very close to violating the strict physical constraint $S + |P| = S_{XX}^{zz} + |S_{XX}^{zz} - S_{YY}^{zz}| \leq 1$ towards the low temperature end of the nematic phase. Furthermore, the results of the *IR* study seem to differ qualitatively from those of the *NMR* study with regards to the molecular mechanism that gives rise to the biaxial order. Thus the major biaxial order parameter according to the *IR* results is $P$, which is insensitive to molecular biaxiality and should vanish as the system tends to perfect biaxial order, while the in the NMR study the observed biaxial order is attributed primarily to the to order parameter $D$, which reflects directly the molecular biaxiality and tends to its maximum value as the system tends to perfect biaxial order. Additionally, within the later interpretation, it is not straightforward to justify how an essentially uniaxial probe solute molecule in a biaxial medium that is dominated by the order parameter $D$ would give rise to the measurement of rather high values ($\eta \sim 0.8$) for the biaxiality parameter.

Whilst the preceding considerations suggest that the assignment of an orthorhombic biaxial symmetry is far from being established on the basis of the present experimental evidence for the thermotropic nematic phases of the bent-core and the calamitic tetrapode systems, there are several experimental observations on these systems that would in fact disfavour an orthorhombic symmetry. In particular, X-ray diffraction studies on bent-core[7, 62] and on structurally similar compounds to the calamitic tetrapodes, including the side-on monomers and the octapodes[63-64] are indicative of the presence of local biaxial order with tilted layers in the nematic phase. The observed macroscopic biaxial order is argued to result from the mutual alignment, spontaneous or externally induced, of these local biaxial structures.[65] Furthermore, in the cases where a higher order mesophase is obtained at lower temperature for any of these compounds, it is invariably of tilted structure, be it tilted smectic[66] or tilted columnar.[63-64] Naturally, these considerations would support the likelihood of a monoclinic, $C_{2h}$, or a triclinic, $C_i$, symmetry. However, the monoclinic symmetry for the biaxial nematic phase of both types of compound is favored over the triclinic symmetry by (i) the presence of at least one plane of symmetry in both the bent-core and the tetrapode molecules (the symmetry being understood in the statistical sense, due to the large number of conformations exhibited by both categories of molecules) in conjunction with (ii) the fact that in all the cases where a more ordered mesophase is obtained on lowering the temperature, this mesophase does have a plane of symmetry.

## VII. CONCLUSIONS

The possible point group symmetries of an a-chiral, a-polar biaxial nematic phase are three: $D_{2h}$ (orthorhombic), $C_{2h}$ (monoclinic) and $C_i$ (triclinic). Our critical review of the presently available experimental results on phase biaxiality in bent-core and in calamitic tetrapode thermotropic nematics suggests that (i) the hitherto routinely assumed, both in theory and in the interpretation of measurements, orthorhombic symmetry may very well not be applicable to these systems and (ii) there are strong indications in support of the monoclinic symmetry. We have identified possible manifestations of the different symmetries on the alignment of biaxial nematics and showed that they could have profound implications on the detection of phase biaxiality in nematics, on its consistent quantification and on its exploitation in electro-optic device applications. We have presented in some detail the methodology for the differentiation between the three possible phase symmetries with two of the most powerful and commonly used experimental techniques for the measurement of phase biaxiality, namely $H^2$-*NMR* and *IR* absorbance spectroscopy. Clearly, for the complete and consistent specification of biaxial order in nematics it is necessary to measure not only the biaxiality parameters $\eta$ associated with the various macroscopic second-rank-tensor properties of the material but also the relative orientations of the principal axes frames of these tensor properties.

**Appendix I. Invariance equations and tensor component relations**

In this appendix we summarize the relations between the components of the $G_{AB}^{(i)}$ tensor in the two principal axes frames *(PAF)* introduced in section IV, namely the $G^{(i)} - PAF$ and the $\chi^m - PAF$. It is recalled that the assignment of the axes $X_i, Y_i, Z_i$ and $X_M, Y_M, Z_M$ of these frames corresponds to ascending principal values of the orientationally averaged field gradient tensor ($\left|G_{X_i X_i}^{(i)}\right| \leq \left|G_{Y_i Y_i}^{(i)}\right| \leq \left|G_{Z_i Z_i}^{(i)}\right|$) and of the magnetic susceptibility ($\chi_{X_M X_M}^m \leq \chi_{Y_M Y_M}^m \leq \chi_{Z_M Z_M}^m$) respectively.

Using the second and third order rotational invariants $G_{AB}^{(i)} G_{AB}^{(i)}$ and $G_{AB}^{(i)} G_{BC}^{(i)} G_{CA}^{(i)}$, the following general relations can be obtained between the parameters $S^{(i)}$ and $\eta^{(i)}$ of Eqs. (5) and (6), the apparent ones $S_M^{(i)}$, $\eta_M^{(i)}$ of Eqs. (15) and (16) and the off diagonal components $G_{Z_M Y_M}^{(i)}, G_{Z_M X_M}^{(i)}, G_{Y_M X_M}^{(i)}$ of the $G_{AB}^{(i)}$ in the $\chi^m - PAF$:

$$\left(S^{(i)}\right)^2 \left(3 + \left(\eta^{(i)}\right)^2\right) = \left(S_M^{(i)}\right)^2 \left(3 + \left(\eta_M^{(i)}\right)^2\right) + 4\left[\left(G_{X_M Y_M}^{(i)}\right)^2 + \left(G_{Y_M Z_M}^{(i)}\right)^2 + \left(G_{Z_M X_M}^{(i)}\right)^2\right] \quad (AI.1)$$

and

$$\left(S^{(i)}\right)^3 \left(1 - \left(\eta^{(i)}\right)^2\right) = \left(S_M^{(i)}\right)^3 \left(1 - \left(\eta_M^{(i)}\right)^2\right) + 8 G_{X_M Y_M}^{(i)} G_{Y_M Z_M}^{(i)} G_{Z_M X_M}^{(i)}$$

$$+ 2 S_M^{(i)} \left[\left(1 + \eta_M^{(i)}\right)\left(G_{Z_M X_M}^{(i)}\right)^2 + \left(1 - \eta_M^{(i)}\right)\left(G_{Y_M Z_M}^{(i)}\right)^2 - 2\left(G_{X_M Y_M}^{(i)}\right)^2\right]$$

(AI.2)

If the phase is known to be orthorhombic, in which case all the off-diagonal components vanish by symmetry, these equations yield the following solutions for the measurable (apparent) parameters $S_M^{(i)}$, $\eta_M^{(i)}$ in terms of $S^{(i)}$ and $\eta^{(i)}$:

$$S_M^{(i)} = S^{(i)} \, ; \, \eta_M^{(i)} = \pm \eta^{(i)}$$

$$S_M^{(i)} = \frac{S^{(i)}}{2}\left(\pm\eta^{(i)} - 1\right) \, ; \, \eta_M^{(i)} = \frac{3 \pm \eta^{(i)}}{1 \mp \eta^{(i)}} \qquad . \qquad (AI.3)$$

$$S_M^{(i)} = \frac{S^{(i)}}{2}\left(\pm\eta^{(i)} - 1\right) \, ; \, \eta_M^{(i)} = -\frac{3 \pm \eta^{(i)}}{1 \mp \eta^{(i)}}$$

Each of these six solutions corresponds to one of the six distinct ways of matching the $X_M, Y_M, Z_M$ with the $X_i, Y_i, Z_i$ axes.

The respective relations in the case of monoclinic biaxial symmetry depend on which one of the $X_M, Y_M, Z_M$ axes and of the $X_i, Y_i, Z_i$ axes coincide with the single two-fold symmetry axis of the phase and involve the angle by which one frame is rotated relative to the other about the common symmetry axis. The relations for the nine different identifications of axis pairs with the symmetry axis are listed in Table I.

**Table I.** Relations between the apparent parameters $S_M^{(i)}$, $\eta_M^{(i)}$ of Eqs. (15) and (16) and the true parameters $S^{(i)}$, $\eta^{(i)}$ of Eqs. (5) and (6), in a biaxial phase of $C_{2h}$ symmetry. Each one of the 9 cells of the table corresponds to a distinct identifications of the two-fold axis of the phase with the indicated pair of axes from the principal frames $X_i, Y_i, Z_i$ (the $G^{(i)} - PAF$) and $X_M, Y_M, Z_M$ (the $\chi^m - PAF$). The angle $g$ of relative rotation of the two frames about their common symmetry axis is identified with the angle $u^{(A_M, B_i)}$ formed by the pair of principal axes $A_M, B_i$ and is expressed in terms of the tensor components in the $\chi^m - PAF$ for each of the 9 possibilities. For notational simplicity the site superscript *(i)* is omitted from the expressions in this table.

| Symmetry axis | $X_i$ | $Y_i$ | $Z_i$ |
|---|---|---|---|
| $X_M$ | $S_M = S\left[\left(\frac{3}{2}\cos^2 g - \frac{1}{2}\right) - \frac{\eta}{2}\sin^2 g\right]$ <br> $\eta_M = -\frac{3(1-\cos 2g) - \eta(3+\cos 2g)}{(3\cos 2g + 1) - \eta(1-\cos 2g)}$ <br> $\tan 2g = \frac{2G_{Z_M Y_M}}{G_{Z_M Z_M} - G_{Y_M Y_M}}$ <br> $g = u^{(Z_M, Z_i)} = u^{(Y_M, Y_i)}$ | $S_M = S\left[\left(\frac{3}{2}\cos^2 g - \frac{1}{2}\right) + \frac{\eta}{2}\sin^2 g\right]$ <br> $\eta_M = -\frac{3(1-\cos 2g) + \eta(3+\cos 2g)}{(3\cos 2g + 1) + \eta(1-\cos 2g)}$ <br> $\tan 2g = \frac{-2G_{Z_M Y_M}}{G_{Z_M Z_M} - G_{Y_M Y_M}}$ <br> $g = u^{(Z_M, Z_i)}$ | $S_M = -\frac{S}{2}(1 - \eta \cos 2g)$ <br> $\eta_M = -\frac{3 + \eta \cos 2g}{1 - \eta \cos 2g}$ <br> $\tan 2g = \frac{2G_{Z_M Y_M}}{G_{Z_M Z_M} - G_{Y_M Y_M}}$ <br> $g = u^{(Y_M, Y_i)}$ |
| $Y_M$ | $S_M = S\left[\left(\frac{3}{2}\cos^2 g - \frac{1}{2}\right) - \frac{\eta}{2}\sin^2 g\right]$ <br> $\eta_M = \frac{3(1-\cos 2g) - \eta(3+\cos 2g)}{(3\cos 2g + 1) - \eta(1-\cos 2g)}$ <br> $\tan 2g = \frac{-2G_{Z_M X_M}}{G_{Z_M Z_M} - G_{X_M X_M}}$ <br> $g = u^{(Z_M, Z_i)}$ | $S_M = S\left[\left(\frac{3}{2}\cos^2 g - \frac{1}{2}\right) + \frac{\eta}{2}\sin^2 g\right]$ <br> $\eta_M = -\frac{3(1-\cos 2g) + \eta(3+\cos 2g)}{(3\cos 2g + 1) + \eta(1-\cos 2g)}$ <br> $\tan 2g = \frac{2G_{Z_M X_M}}{G_{Z_M Z_M} - G_{X_M X_M}}$ <br> $g = u^{(Z_M, Z_i)} = u^{(X_M, X_i)}$ | $S_M = -\frac{S}{2}(1 + \eta \cos 2g)$ <br> $\eta_M = \frac{3 - \eta \cos 2g}{1 + \eta \cos 2g}$ <br> $\tan 2g = \frac{-2G_{Z_M X_M}}{G_{Z_M Z_M} - G_{X_M X_M}}$ <br> $g = u^{(X_M, X_i)}$ |
| $Z_M$ | $S_M = -\frac{S}{2}(1-\eta)$ <br> $\eta_M = -\frac{3+\eta}{1-\eta}\cos 2g$ <br> $\tan 2g = \frac{-2G_{X_M Y_M}}{G_{X_M X_M} - G_{Y_M Y_M}}$ <br> $g = u^{(Y_M, Y_i)}$ | $S_M = -\frac{S}{2}(1+\eta)$ <br> $\eta_M = \frac{3-\eta}{1+\eta}\cos 2g$ <br> $\tan 2g = \frac{2G_{X_M Y_M}}{G_{X_M X_M} - G_{Y_M Y_M}}$ <br> $g = u^{(X_M, X_i)}$ | $S_M = S$ <br> $\eta_M = \eta \cos 2g$ <br> $\tan 2g = \frac{-2G_{X_M Y_M}}{G_{X_M X_M} - G_{Y_M Y_M}}$ <br> $g = u^{(Y_M, Y_i)} = u^{(X_M, X_i)}$ |

**Appendix II. Calculation of Deuterium NMR spectral patterns from orientationally distributed biaxial nematic samples**

**AII.a. Outline of the deuterium NMR spinning sample method.** Briefly, this method involves (i) using the static measurement to extract the $G^{(i)}_{Z_M Z_M}$ component from the value of the splitting of the magnetically aligned sample according to Eq. (15) and (ii) analyzing the spinning sample spectral pattern in order to extract the values of the biaxial, $\left(G^{(i)}_{X_M X_M} - G^{(i)}_{Y_M Y_M}\right)$, and the off-diagonal, $G^{(i)}_{Z_M Y_M}$, components. In the simplest implementation of such analysis, the spinning of the sample is assumed to generate a uniform planar distribution of $Z_M$ in the plane perpendicular to the spinning axis (and therefore containing the magnetic field), with the $X_M$ axis remaining perpendicular to the magnetic field (and therefore parallel to the spinning axis) for all the distributed orientations of $Z_M$ relative to the magnetic field. Under these assumptions, $\cos\phi_M = 0$ in Eq. (12) and therefore the components $G^{(i)}_{Z_M X_M}, G^{(i)}_{Y_M X_M}$ do not contribute to the spectral pattern. The latter is governed by the following $\theta_M$ dependence of the spinning-sample splittings

$$\delta\nu^{(i)}_S(\theta_M) = \frac{3}{2}\nu^{(i)}_Q\left[E^{(i)} + F^{(i)}\cos 2\left(\theta_M - g^{(i)}\right)\right] \tag{AII.1}$$

where

$$E^{(i)} = \frac{S^{(i)}_M}{4}\left(1 - \eta^{(i)}_M\right)$$

$$F^{(i)} = \sqrt{\left(\frac{S^{(i)}_M}{4}\left(3 + \eta^{(i)}_M\right)\right)^2 + \left(G^{(i)}_{Y_M Z_M}\right)^2} \tag{AII.2}$$

$$\tan 2g^{(i)} = \frac{4 G^{(i)}_{Y_M Z_M}}{S^{(i)}_M\left(3 + \eta^{(i)}_M\right)}$$

and the line shape generated by the spinning sample spectra, in the absence of line broadening, is described, to within an overall scale constant, by the function[61]

$$L^{(i)}(\nu) = \left(\left(F^{(i)}\right)^2 - \left(\frac{4\nu}{3\nu^{(i)}_Q} - E^{(i)}\right)^2\right)^{-1/2} + \left(\left(F^{(i)}\right)^2 - \left(\frac{4\nu}{3\nu^{(i)}_Q} + E^{(i)}\right)^2\right)^{-1/2}. \tag{AII.3}$$

The peaks of this symmetric line shape at $\nu = \pm 3\nu^{(i)}_Q\left[E^{(i)} \pm F^{(i)}\right]/4$ provide the values of the parameters $E^{(i)}$ and $F^{(i)}$. Then the angle $g^{(i)}$ can be obtained from the splitting of the aligned sample using Eq. (AII.1) for $\theta_M = 0$ (see Fig. AII.1). Having evaluated $E^{(i)}$, $F^{(i)}$ and $g^{(i)}$, the values of the apparent parameters $S^{(i)}_M$ and $\eta^{(i)}_M$ and of the off diagonal component $G^{(i)}_{Y_M Z_M}$ are obtained though Eqs. (AII.2).

For the case of $C_{2h}$ symmetry and $X_M$ coinciding with the symmetry axis, Eqs. (AI.1) and (AI.2), combined with Eq. (AII.2) lead to the following relations, from which $S^{(i)}$ and $\eta^{(i)}$ can be directly evaluated from $E^{(i)}$ and $F^{(i)}$,

$$\left(\frac{S^{(i)}}{2}\right)^2 \left(3+\left(\eta^{(i)}\right)^2\right) = \left(F^{(i)}\right)^2 + 3\left(E^{(i)}\right)^2 \quad ,$$

$$\left(\frac{S^{(i)}}{2}\right)^3 \left(1-\left(\eta^{(i)}\right)^2\right) = E^{(i)}\left(\left(F^{(i)}\right)^2 - \left(E^{(i)}\right)^2\right)$$

(AII.4)

It should be noted that these relations do not involve the angle $g^{(i)}$; the latter is to be evaluated independently from the aligned sample splitting through Eq. (AII.2).

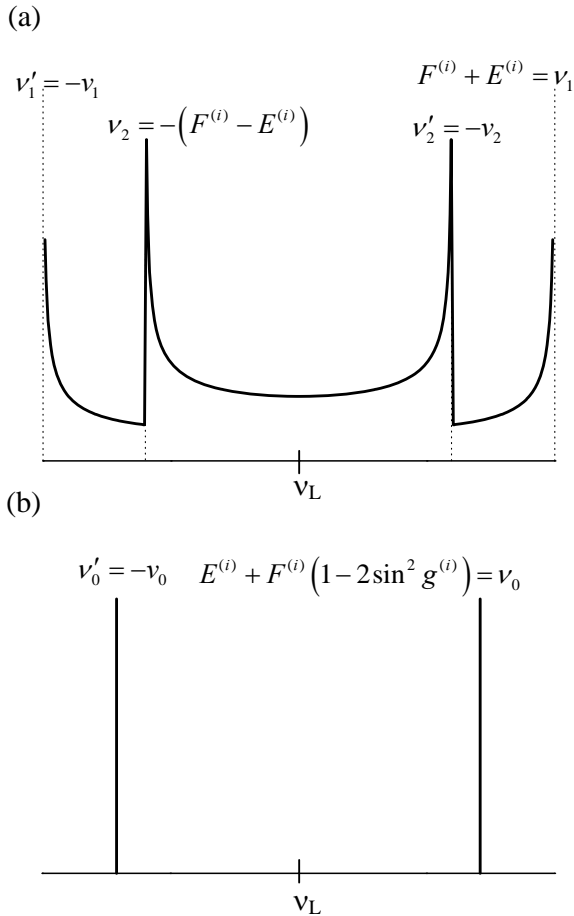

**FIG. AII.1.** (a) Spinning sample line-shape calculated according to Eq. (AII.3) for a deuteriated site in the case where $X_i \parallel X_M$ and using the value $\eta^{(i)} = 0.2$ for the biaxiality parameter. The frequencies are expressed in units of $3\nu_Q^{(i)}/4$. (b) The calculated spectrum for the same deuteriated site obtained from the magnetically aligned sample using the value $25^o$ for the angle $g^{(i)}$ in Eq. (AII.2).

Deviations from the idealized conditions assumed in the derivation of the line-shape in Eq. (AII.3) include (i) possible non planarity and non-uniformity of the distribution of the principal diamagnetic axis $Z_M$ in the plane perpendicular to the spinning axis, (ii) the presence of line broadening and, (iii) most important at high spinning rates, the interference of the sample-spinning frequency with the frequency variable $\nu$ of the line-shape. The details of the spectral analysis taking into account such deviations are considered in Refs. 53, 61 and 67.

**AII.b. Spectra with chemical shift asymmetry (*CSA*).**
For a deuteriated site exhibiting $CSA^{50}$ the frequencies of the quadrupolar spectrum involve, in addition to the average field-gradient tensor $G_{AB}^{(i)}$, the *CSA* tensor $C_{AB}^{(i)}$ whose principal axis frame, the $C^{(i)} - PAF$, need not necessarily coincide with the $G^{(i)} - PAF$. In this case Eq. (12), describing the splitting of a symmetric quadrupolar spectrum, should be replaced by a more general expression describing the two frequencies $\nu^{(i)\pm}$ of the asymmetric spectrum. For $X_M$ directed perpendicular to the magnetic field (i.e $\cos\phi_M = 0$) these frequencies have the following dependence on the angle $\theta_M$:

$$\nu^{(i)\pm}(\theta_M, \cos\phi_M = 0) = \nu_L(1 + \sigma_{iso}^{(i)})$$
$$+\nu_L \sigma_2^{(i)} \left[ \left(\frac{3}{2}\cos^2\theta_M - \frac{1}{2}\right) C_{Z_M Z_M}^{(i)} - \frac{1}{2}\sin^2\theta_M \left(C_{X_M X_M}^{(i)} - C_{Y_M Y_M}^{(i)}\right) + \sin 2\theta_M \sin\phi_M C_{Z_M Y_M}^{(i)} \right]$$
$$\pm \frac{3}{4}\nu_Q^{(i)} \left[ \left(\frac{3}{2}\cos^2\theta_M - \frac{1}{2}\right) G_{Z_M Z_M}^{(i)} - \frac{1}{2}\sin^2\theta_M \cos 2\phi_M \left(G_{X_M X_M}^{(i)} - G_{Y_M Y_M}^{(i)}\right) + \sin 2\theta_M \sin\phi_M G_{Z_M Y_M}^{(i)} \right]$$

(AII.5)

Here $\nu_L$ is the Larmor frequency of the deuteron, and $\sigma_{iso}^{(i)}, \sigma_2^{(i)}$ denote respectively the isotropic and anisotropic (rank 2) *CSA* coupling constants[50] for the deuterated site *i*. Defining the parameter $\lambda^{(i)} \equiv \dfrac{4\nu_L \sigma_2^{(i)}}{3\nu_Q^{(i)}}$ and the tensors

$$G_{A_M B_M}^{(i)\pm} \equiv G_{A_M B_M}^{(i)} \pm \lambda^{(i)} C_{A_M B_M}^{(i)} \quad, \tag{AII.6}$$

the spectral frequencies of Eq. (AII.5) can be written as

$$\nu^{(i)\pm}(\theta_M, \cos\phi_M = 0) = \nu_L(1 + \sigma_{iso}^{(i)}) \pm \frac{3}{4}\nu_Q^{(i)} \left[ E^{(i)\pm} + F^{(i)\pm} \cos 2(\theta_M - g^{(i)\pm}) \right] \quad, \tag{AII.7}$$

with

$$E^{(i)\pm} = \left(G_{Z_M Z_M}^{(i)\pm} - G_{X_M X_M}^{(i)\pm} + G_{Y_M Y_M}^{(i)\pm}\right)/4$$

$$F^{(i)\pm} = \sqrt{\left(\left(3G_{Z_M Z_M}^{(i)\pm} + G_{X_M X_M}^{(i)\pm} - G_{Y_M Y_M}^{(i)\pm}\right)/4\right)^2 + \left(G_{Y_M Z_M}^{(i)\pm}\right)^2} \quad, \tag{AII.8}$$

and

$$\sin^2 g^{(i)\pm} = \frac{E^{(i)\pm} + F^{(i)\pm} \pm \dfrac{4}{3\nu_Q^{(i)}} \left(\nu_L(1 + \sigma_{iso}^{(i)}) - \nu^{(i)\pm}(\theta_M = 0)\right)}{2F^{(i)\pm}} \quad. \tag{AII.9}$$

The two lines will generate a spectral pattern which, with the frequency origin placed at $\nu_L(1 + \sigma_{iso}^{(i)})$, has the expression

$$L^{(i)}(\nu) = \left(\left(F^{(i)+}\right)^2 - \left(\frac{4\nu}{3\nu_Q^{(i)}} - E^{(i)+}\right)^2\right)^{-1/2} + \left(\left(F^{(i)-}\right)^2 - \left(\frac{4\nu}{3\nu_Q^{(i)}} + E^{(i)-}\right)^2\right)^{-1/2}. \tag{AII.10}$$

The peaks of the line shape provide the values of the four parameters $E^{(i)\pm}, F^{(i)\pm}$ (in units of $\frac{3}{4}\nu_Q^{(i)}$) which, combined with the values of the aligned-sample spectral frequencies as in (AII.9), lead to the evaluation of the angles $g^{(i)\pm}$. Then, with the help of Eqs. (AII.6) and (AII.7) the tensor components $G_{Z_M Z_M}^{(i)}, \left(G_{X_M X_M}^{(i)} - G_{Y_M Y_M}^{(i)}\right), G_{Y_M Z_M}^{(i)}$ and $C_{Z_M Z_M}^{(i)}, \left(C_{X_M X_M}^{(i)} - C_{Y_M Y_M}^{(i)}\right), C_{Y_M Z_M}^{(i)}$ can be evaluated. In the case of $C_{2h}$ (or higher) phase symmetry, and with $X_M$ coinciding with the symmetry axis, this constitutes a complete determination of the tensors $G_{AB}^{(i)}$ and $C_{AB}^{(i)}$, and therefore of the deviation angles of their respective principal axis frames from the $\chi^m - PAF$.

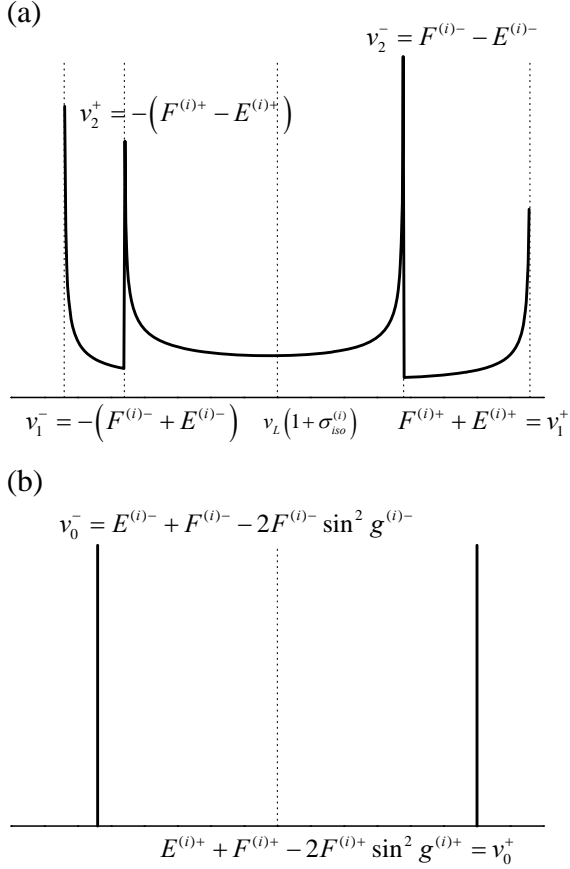

**FIG. AII.2.** (a) Spinning sample line-shape calculated according to Eq. (AII.10) for a deuteriated site in the presence of chemical shift asymmetry *(CSA)* and for $X_i \parallel X_M$. The input parameter values of the calculation are: $\lambda^{(i)} = 0.1$ for the *CSA* relative strength parameter; $S^{(i)} = 0.7, \eta^{(i)} = 0.2, g^{(i)} = 20^o$ for the quadrupolar *G*-tensor parameters; $S^{(i)} = 0.7, \eta^{(i)} = 0.4, g^{(i)} = 35^o$ for the *CSA* *C*-tensor parameters. The frequencies are expressed in units of $3\nu_Q^{(i)}/4$. (b) The calculated spectrum for the same deuteriated site obtained from the magnetically aligned sample.

**Acknowledgement**
The research leading to these results has received funding from the European Community's Seventh Framework Programme (FP7/2007-2013) under Grant Agreement n° 216025-BIND (Biaxial Nematic Devices).